\documentclass[aps,prb,twocolumn,preprintnumbers]{revtex4}
\usepackage{graphics}
\usepackage{graphicx}
\usepackage{dcolumn} 
\usepackage{bm}
\usepackage{color}
\usepackage{epsfig}
\usepackage{amsmath}
\begin{document}
\title{Cubic H$_3$S around 200 GPa: an atomic hydrogen superconductor stabilized by sulfur 
}
\author{D. A. Papaconstantopoulos$^1$}
\author{B. M. Klein$^2$}
\author{M. J. Mehl$^3$}
\author{W. E. Pickett$^{2}$}
\email{pickett@physics.ucdavis.edu}
\affiliation{
$^1$Computational Materials Science Center, George Mason University,
  Fairfax, VA 22030, USA\\
$^2$Department of Physics, University of California Davis, Davis, CA 95616, USA\\
$^3$Center for Computational Materials Science, Naval Research Laboratory, Washington DC 20375, USA
}
\date{\today}
\begin{abstract}
The multiple scattering-based theory of Gaspari and Gyorffy for the
electron-ion matrix element in close packed metals is applied to
$Im{\bar 3}m$ H$_3$S, which has been predicted by Duan {\it et al.}
and Bernstein {\it et al} to be the stable phase at this
stoichiometry around 190 GPa, thus is the leading candidate to be
the phase observed to superconduct at 190K by Drozdov, Eremets, and
Troyan. The nearly perfect separation of vibrational modes into
those of S and of H character provides a simplification that enables
identification of contributions of the two atoms separately. The
picture that arises is basically that of superconducting atomic H
stabilized by strong covalent mixing with S $p$ and $d$ character.
The reported isotope shift is much larger than the theoretical one,
suggesting there is large anharmonicity in the H vibrations. Given
the relative unimportance of sulfur, hydrides of lighter atoms at
similarly high pressures may also lead to high temperature
superconductivity.
\end{abstract}
\maketitle
\vskip 0.7in

\section{Background}

The report by Drozdov, Eremets, and Troyan\cite{Drozdov} (DET) of
superconductivity up to T$_c$=190 K in H$_2$S compressed to the 200
GPa regime has reignited excitement in the possibility of achieving
room temperature superconductivity. This report builds on previous
success of pressure enhancement of T$_c$ in a variety of types of
materials: from 134K to 164K in the cuprate
Hg2223,\cite{Hg2223,Hg2223v2} from zero to 20-25 K in the simple
metals Li, Ca, and
Y,[\onlinecite{sch1,Li1,Li2,DAPli,calcium,yttrium,CaY}] and from
zero to 14K in the insulator silicon.\cite{silicon} An anticipated
major factor is the increase in the phonon energy scale with
compression, since it sets the temperature scale for T$_c$, as
pointed out early on\cite{ashcroft} and reviewed more
recently\cite{ashcroft2} by Ashcroft in predicting possible room
temperature superconductivity in metallic hydrogen.

The newly reported high values of T$_c$ appear to confirm
theoretical predictions that predated the experiment. Applying
particle swarm crystal structure search techniques founded on
density functional theory, Li {\it et al.} predicted\cite{Li}
candidate stable crystal structures of H$_2$S up to 220 GPa. These
predictions were followed by linear response calculations of the
phonon spectrum $\omega_Q$, electron-phonon matrix elements, and
finally the Eliashberg spectral function $\alpha^2F(\omega)$, from
which T$_c$ can be calculated, depending only mildly on the presumed
value of the retarded Coulomb repulsion $\mu^*$ = 0.10-0.13. For
pressures of 140-180 GPa, they obtained an electron-phonon coupling
strength $\lambda$=1.0-1.2, an Allen-Dynes characteristic
frequency\cite{alldyn} $\omega_{log}$$\sim$1000K, and T$_c$ of 40K
at 140 GPa and peaking at 80 K at 160 GPa. While 80K is well short
of the reported T$_c$=190K, the result is convincing that very high
T$_c$ is predicted in H$_2$S at high pressure.

The sister stoichiometry H$_3$S has been explored in very similar
fashion by Duan {\it et al.}\cite{Duan} Predicting structures to
more than 200 GPa, their linear response results for $Im{\bar 3}m$
H$_3$S led to very large calculated values of electron-phonon
coupling strength $\lambda$=2.0-2.2, frequency scales
$\omega_{log}\sim$1300K, and values of T$_c$ up to 200K. In the
calculations of Li {\it et al.} and Duan {\it et al.}, the high
values of $\omega_{log}$ are expected from the anticipated increase
of force constants as volume is decreased. The large values of
$\lambda$, a factor of two or more over most other very good
superconductors (including MgB$_2$), imply that the electronic
matrix elements are substantially larger than in nearly all known
conventional superconductors.

In this report we use Gaspari-Gyorffy (GG) theory\cite{GG} to
provide insight into why electron-ion matrix elements vary, and
evidently increase strongly, with pressure. Such understanding is
necessary not only to interpret the results described above, but
also to provide essential clues how to increase matrix elements, and
$\lambda$, at lower or possibly ambient pressure.  Interestingly,
shortly after the formulation of this theory, two of the present
authors applied it to predict T$_c$$\sim$250K in metallic hydrogen
at a few Mbar pressure.\cite{hydrogen}

GG theory\cite{GG} builds on the earlier observation of
Hopfield\cite{Hopfield} that electron scattering off (moving) ions
has strong local character. First, metallic screening means the
Thomas-Fermi screening length is short, of the order of 1~\AA, and
very weakly dependent on carrier density (more correctly, the Fermi
level density of states [DOS] N(E$_F$)). GG employed a multiple
scattering Green's function formalism that facilitated three
simplifications. The first is that the potential is spherical (very
good approximation) and is negligible beyond the atomic sphere; the
second is that the linear change in potential of a displaced ion can
be approximated by a rigid shift of the atomic potential.  Thirdly,
the direction dependence of the wavefunction coefficients is
averaged out, thereby neglecting any special influence of Fermi
surface shape. The H$_3$S Fermi surface\cite{bernstein} is large and
multisheeted, minimizing the likelihood of Fermi surface effects.
It is possible that these approximations may improve with reduction
in volume, in any case these approximations should not degrade as
the system become denser.  The bands shown by Duan {\it et
  al.}\cite{Duan} and Bernstein {\it et al.}\cite{bernstein} for the
$Im{\bar 3}m$ structure of H$_3$S that we discuss show much free
electron, spherical character in the lower 75\% of the occupied
bands, though less so around the Fermi energy where S $3p$ - H $1s$
hybridization produces structure in the DOS.

\begin{figure}[tbp]
{\resizebox{5.0cm}{5.0cm}{\includegraphics{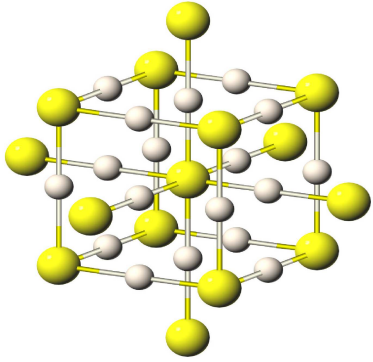}}}
\caption{(Color online) The $Im{\bar 3}m$ crystal structure of
  H$_3$S, illustrating the two interlaced ReO$_3$ sublattices. Large
  sphere is S, small sphere is H.  }
\label{Structure}
\end{figure}

\section{Theory and Results}

The coupling strength $\lambda$, and the frequency weighting
spectrum $g(\omega)$ normalized to unity, are given in Eliashberg
theory\cite{SSW} by
\begin{eqnarray}
\lambda = 2\int \frac{\alpha^2F(\omega)}{\omega}d\omega;
 ~~~~~~ g(\omega)=\frac{2\alpha^2F(\omega)}{\lambda \omega},
\label{eqn1}
\end{eqnarray}
where $\alpha^2F$ is the Eliashberg electron-phonon spectral
function that governs many superconducting properties.  The
calculations by Li {\it et al.} for H$_2$S and Duan {\it et al.}
for H$_3$S demonstrate that the lower range of phonon frequencies
(the acoustic modes) have negligible H character, while the optic
modes above the gap at 20-25 THz have negligible S character, making
it an ideal platform for applying the GG expression to the atoms
{\it separately}.

Thus $\lambda = 3\lambda_H +\lambda_S$, where the latter arise from
the integral over the low frequency S modes, the former from the
nine higher frequency H branches.  In this case the GG expression,
given originally for an elemental solid, can be applied to the S and
H spheres separately.\cite{klein76} Each atomic ($j$) coupling
constant $\lambda_{j}$ is given by the integral over the appropriate
frequency region, leading to
\begin{eqnarray}
\lambda_{j} = \frac{N(E_F) <I_j^2>}{M_j \omega^2_j}
             \equiv \frac{\eta_j}{M_j \omega^2_j}.
\end{eqnarray}
The averaged matrix elements $<$$I^2$$>$ obtained from GG theory are
discussed below.

The separation of mode character also allows a simple estimate of
the total frequency moments that enter the Allen-Dynes (AD) equation
for T$_c$, through the weight function
\begin{eqnarray}
g(\omega) = [\lambda_S g_S(\omega) + 3\lambda_H g_H(\omega)]/(\lambda_S + 3\lambda_H),
\end{eqnarray}
where the partial $g_{j}$ functions, defined analogously to that of
AD (Eq. \ref{eqn1}) are separately normalized to unity.

\begin{figure}[tbp]
{\resizebox{8.5cm}{5.5cm}{\includegraphics{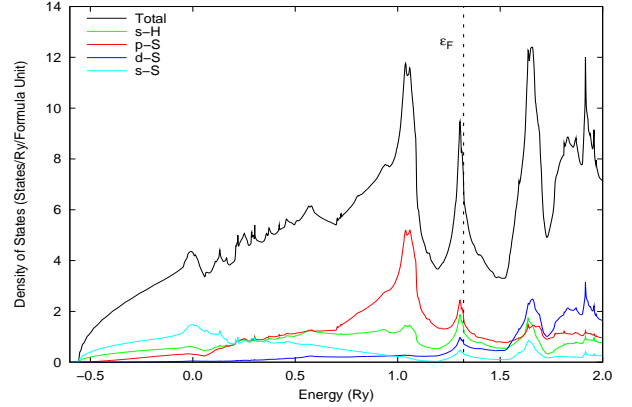}}}
\caption{(Color online) Total and orbital-projected densities of
  states of H$_3$S at 210 GPa ($a$=5.6 a.u.). Note the position of
  E$_F$ within the sharp and narrow peak, and that the H $1s$
  contribution is that from all three H atoms.  }
\label{DOS1}
\end{figure}

Both H$_2$S and H$_3$S having been shown\cite{Li,Duan} to have
strong electron-phonon coupling at high pressure.  Bernstein {\it et
  al.}\cite{bernstein} have provided convincing evidence that H$_2$S
is unstable to decomposition into bcc H$_3$S and sulfur, and that
competing stoichiometries are unlikely. This result confirms the
suggestion of DET, who reported sulfur formation in their samples.
Because of this evidence on the most likely superconducting phase,
we focus on $Im{\bar 3}m$ H$_3$S with its simple $bcc$ structure
based in two interlaced ReO$_3$ structure sublattices.  From
$\alpha^2F$ of Duan {\it et al.}, we simplify with a constant
$\alpha^2F_j$ (constant for each atom species $j$) with frequency
ranges (in kelvin) of [430,820] for S and [1250,2500] for H. Results
are insensitive to these limits, depending mostly on the mean
frequencies and the separation of $\lambda$ into S and H
contributions.  The resulting frequency moments $\omega_{log},
\omega_1$, and $\omega_2$ and associated data, for insight into
separate S and H contribution, for use in the AD
equation,\cite{alldyn} and to compare with results below from GG
theory, are presented in Table~\ref{DuanTable}.

With $\mu^*$=0.15, T$_c$=234 K results; the difference from the
value quoted by Duan {\it et al.} might be due to neglect of the
strong coupling factor $f_1$=1.13 factor in the AD equation, which
amounts to a 26 K increase, or partially to our constant $\alpha^2F$
modeling. Neglecting the contribution from the S modes, $\lambda$ is
decreased from 2.2 to 1.5 but $\omega_{log}$ increases from 1500 K
to 1770 K, and T$_c$ decreases by only 19 K to 215 K. The sulfur
contribution to T$_c$ is 8\%; H$_3$S is basically an atomic hydrogen
high temperature superconductor.  Bernstein {\it et al.} also
suggested that S vibrations are not essential for the high
T$_c$. The H isotope effect can also be obtained.  $\lambda$ is
unaffected by masses; $M\omega^2$ is a function of the force
constants alone, so frequencies, specifically $\omega^H_{log}$
decrease as the square root of the mass. The resulting critical
temperature is reduced to $\sim$170 K, slightly more than 234
K/$\sqrt{2}$ because the small S contribution remains unchanged. The
experimental value of DET\cite{Drozdov} is 90 K; the most likely
cause of this discrepancy is strong anharmonicity of the H optic
modes.

The second moment frequency at 200 GPa of S is $\omega^S_2$=615 K,
while that of H is $\omega^H_2$=1840 K, thus with the atomic masses
of 32 and 1 a.m.u respectively, the denominator $M\omega_2^2$ is
32/9 =3.5 larger for S.  The consequence is that a given
contribution to $\eta_H$ is 3.5$\times$3$\sim$10 times more
effective in increasing $\lambda$ than the same contribution to
$\eta_S$ (though in practice there is no clear method of effecting
such a tradeoff).

\begin{table}[!ht]
\caption{Electron-phonon coupling data for H$_3$S, obtained from
  modeling the results of Duan {\it et al.} with a constant
  $\alpha^2F$ model. Frequencies are in Kelvin.}
\begin{ruledtabular}
\begin{tabular}{lcccc}
\multicolumn{1}{l}{} &
\multicolumn{1}{c}{S} &
\multicolumn{1}{c}{H} &
\multicolumn{1}{c}{H$_3$S} & \\
\hline
$\omega_{log}$& 595  & 1770 & 1500 \\
$\omega_1$    & 605  & 1800 & 1530 \\
$\omega_2$    & 615  & 1840 & 1560 \\
$M\omega^2_2$ (eV/\AA$^2$) & 9.3  & 2.6  & --\\
$\eta$        (eV/\AA$^2$) & 4.7 & 1.48 & --\\
$\lambda$     & 0.5      & 1.7/3    & 2.2
\label{DuanTable}
\end{tabular}
\end{ruledtabular}
\end{table}

In terms of the phase shifts $\delta_{\ell,j}$ for the $j$-th atom
for orbital channel $\ell$, the square electron-ion matrix element
averaged over the Fermi surface can be written in the simple
form\cite{GG} as
\begin{eqnarray}
  \mbox{$<$$I^2_{j}$$>$}=\frac{E_F}{\pi^2} \frac{1}{N(E_F)^2}
  \sum_{\ell=0}^{2} 2({\ell}+1) \sin^2(\delta_{\ell}-\delta_{\ell
    +1}) \nu_{\ell}\nu_{\ell +1},
\end{eqnarray}
where $\nu_{\ell}=N_{\ell}(E_F)/N^{(1)}_{\ell}$ is the ratio of the
${\ell}$-th partial DOS to $N^{(1)}$, the single scatterer DOS, for
the given atomic potential in a homogeneous system. $<$$I^2$$>$ is
independent of N(E$_F$) since it can equally well be
expressed\cite{wep} in terms of the fractions $N_{\ell}(E_F)/N(E_F)$
which usually do not reflect the van Hove singularities of either
one. The calculated DOS at 210 GPa is shown in Fig.~\ref{DOS1}. The
Fermi level falls at a sharp and narrow peak; calculations at other
volumes indicate this is a persistent occurrence.  $<$$I^2_{j}$$>$
will tend to be maximized in the cases where ``neighboring''
channels $\ell,\ell+1$ have a large difference in phase shifts but
ratios $\nu_{\ell}$ that are as large as possible. From the GG
expression, for each atom $\eta_j=N(E_F)$$<$$I_j^2$$>$, and the
latter factor involves the $\sin^2(\delta_l-\delta_{l+1})$ factor
and products of PDOS ratios $\nu_l \nu_{l+1}$.  $M_j\omega_j^2$ can
be expressed in terms of the ionic force constants, independent of
$M_j$, (which we return to below) so that any isotope effect
different from $M^{-1/2}$ will arise from factors beyond $\lambda$
(primarily anharmonicity).

The calculations have been carried out with two all-electron
linearized augmented plane wave (LAPW) codes, one developed at
NRL,\cite{nrllapw} and also ELK.\cite{elk} The sphere radii were 1.8
a.u. and 1.0 a.u. for S and H respectively, except for the smallest
volume where the S radius was reduced because the sum of the radii
must be no more than $a$/2.

The band structure\cite{Duan,bernstein} consists of four nearly
filled bands, leaving some holes at $\Gamma$ and electrons around
N. In addition, a fifth broad band is roughly half filled. The DOS
is noteworthy: free electron like over 20 eV of the valence band
before strong structure arises in a $\pm$7 eV range centered at
E$_F$, which lies very close (slightly above) the strong and sharp
peak in the DOS. This peak at E$_F$ persists for all pressures from
P=0 to 300 GPa and even above, almost as if E$_F$ were pinned at
this peak, while other features of the DOS evolve.

Table~\ref{SmallTable} shows the Fermi level values of total and
angular momentum components of the electronic densities of states
across a wide range of volumes.  Even though $Im{\bar 3}m$ H$_3$S
may not be stable at lower pressures, we provide results for the
large range P=0-210 GPa to observe the effect of interatomic
distance on the electronic structure and coupling.  While the total
N(E$_F$) shows a weak non-monotonic variation, the $\ell$-components
have a stronger lattice constant(pressure) dependence.  As expected
the sulfur $3p$-like states are the dominant component but a strong
participation of $3d$ character especially at high pressures is
present, hybridizing with the also strong and nearly constant H $1s$
contribution.

\begin{figure}[tbp]
{\resizebox{8.5cm}{5.5cm}{\includegraphics{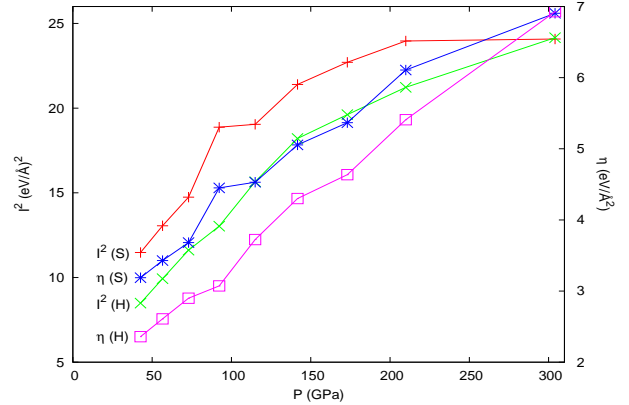}}}
\caption{(Color online) Plots demonstrating the very strong pressure
  increase of $<$$I^2(j)$$>$ (left axis) and $\eta_j$ (right axis)
  for $j$=S and H in $Im{\bar 3}m$ H$_3$S. For H, the $\eta$ plotted
  is the total for 3 H atoms.  }
\label{i2eta}
\end{figure}

Now we discuss the electronic factor $\eta_{j}$.  It should be kept
in mind that the relative importance for T$_c$ of H versus S modes
is not simply $\lambda_j$, but more like $\lambda_{j}\omega_{2,j}$
which is a factor of ten greater for 3H than for S).  For hydrogen,
which dominates the contribution to T$_c$ only the $s-p$ channel is
relevant at all pressures. The phase shift factor $\sin^2(\delta_s -
\delta_p)$ {\it decreases} with pressure. Fig.~\ref{i2eta}
illustrates the factor of two increase in $\eta_H$ from P=100 GPa to
300 GPa. This dramatic increase results from an even larger increase
in the PDOS product $\nu^H_s \nu^H_p$, reflecting transfer of $1s$
character to $2p$ character. In the spherical harmonic expansion of
the atomic wavefunctions this `$2p$' character represent the
expansion of tails of the orbitals on neighboring atoms that gives
rise to the increased H-S hybridization under pressure.

\begin{table}
\caption{Total and angular-momentum decomposed densities of electronic
states at the Fermi level (states/Ry-both spins)
in $Im{\bar 3}m$ H$_3$S. The $\ell$-components
are projections within sphere radii R(S)=1.8 a.u. and R(H)=1.0 a.u.}
\begin{tabular}{c|c|ccc|c}
\hline
$a$ (a.u.)&N(E$_F$)&   S-s &   S-p &   S-d &  H-s/atom \\
\hline
 5.6 &  6.93 &  0.325 &  1.74 &  0.751 &  0.435\\
 5.8 &  6.43 &  0.286 &  1.53 &  0.605 &  0.405\\
 6.0 &  6.47 &  0.274 &  1.46 &  0.528 &  0.415\\
 6.2 &  6.42 &  0.253 &  1.18 &  0.528 &  0.462\\
 6.4 &  6.79 &  0.259 &  1.36 &  0.417 &  0.455\\
 6.6 &  7.15 &  0.255 &  1.38 &  0.375 &  0.482\\
 6.8 &  7.56 &  0.254 &  1.42 &  0.339 &  0.513\\
\hline
\end{tabular}
\label{SmallTable}
\end{table}

\begin{table*}[!]
\caption{Pressure variation of the Hopfield parameter $\eta$
  (eV/\AA$^2$) and electron-ion squared matrix element $<$$I^2$$>$
  (eV$^2$/\AA$^2$), both decomposed into the three channels $sp$,
  $pd$, and $df$ for S, and the H $sp$ channel (given for a single
  atom).}
\begin{tabular}{r|r|cccccc|cc}
\hline
a (a.u.)&P (GPa)&S $\eta_{sp}$&S I$^2_{sp}$&S $\eta_{pd}$&S I$^2_{pd}$&S $\eta_{df}$&S I$^2_{df}$&H $\eta_{sp}$&H I$^2_{sp}$\\
\hline
5.6     & 210   & 0.68        & 2.66       & 5.12        &  20.1      & 0.31        &  1.21      & 1.80        & 7.06 \\
5.8     & 142   & 0.39        & 1.65       & 4.45        &  18.9      & 0.21        &  0.89      & 1.43        & 6.06 \\
6.0     &  92   & 0.24        & 1.00       & 4.14        &  17.4      & 0.15        &  0.63      & 1.24        & 5.21 \\
6.2     &  57   & 0.19        & 0.78       & 4.13        &  17.5      & 0.13        &  0.57      & 1.02        & 4.34 \\
6.4     &  31   & 0.08        & 0.33       & 3.52        &  14.1      & 0.08        &  0.31      & 0.96        & 3.87 \\
6.6     &  13   & 0.05        & 0.18       & 3.37        &  12.7      & 0.06        &  0.21      & 0.87        & 3.31 \\
6.8     &   0   & 0.02        & 0.09       & 3.12        &  11.2      & 0.04        &  0.14      & 0.79        & 2.83 \\
\hline
\end{tabular}
\label{BigTable}
\end{table*}

As noted above from our analysis of the results of Duan {\it et
  al.}, the sulfur contribution is less important. As for H, the
increase in $\eta$ occurs in spite of a {\it decrease} in the
$\sin^2(\delta_p-\delta_d)$ factor, by 30\% from P=100 to P=300 GPa.
Over this pressure range, the PDOS ratio product $\nu^S_p\nu^S_d$
increases by 75\%, giving a net $\eta$ increase by more than 50\%.
In addition, the S $s-p$ and even $d-f$ channels begin to
contribute, reaching 20\% of the total of $<$$I^2$$>$ at 300
GPa. Thus the increase is a composite effect of increase of both $d$
and $s$ character of S, which is reflected also in the growing phase
shifts of these channels.  These transfers of atomic character under
pressure are consistent with general expectations of the evolution
of atomic character under reduction of volume.  Interpolating to 200
GPa to obtain $\eta_S$=5.84 eV/\AA$^2$, $\eta_H$=1.71 eV/\AA$^2$
(which must be multiplied by three), and using the frequency moments
from Table 1, we obtain $\lambda$=2.6, thus T$_c\sim 250 K$
depending somewhat on the chosen value of $\mu^*$=0.10-0.13. This
value is very consistent with the value of T$_c$ (above) obtained
from our modeling of the Duan data.  Though the numbers might vary
for other low-Z hydrides (H$_n$P, H$_n$B, ...) the lack of any
special role of S in these results suggests there should also be a
strong increase in T$_c$ with pressure in other low-Z element
hydrides.

A question of great interest is whether T$_c$ increases further at
higher pressure. The data presented in Table~\ref{BigTable} provides
the pressure dependence of the important quantities entering
$\eta_S$ and $\eta_H$. This data demonstrates that the strongest
contributions arise for sulfur from the $pd$ channel and for
hydrogen from the $sp$ channel. Note that for H one should multiply
by three to account for the three H atoms in the unit cell.

Since $\lambda_H$ dominates the sulfur contribution, we can focus on
the H contribution alone. The total pressure derivative contains
several contributions
\begin{eqnarray}
\frac{d {\rm log} T_c}{dP}&=&\frac{d {\rm log} \omega_{log}}{dP} + \frac{d {\rm log} f_1}{dP}
                     + \frac{d {\rm log} E(\lambda)}{dP}  \nonumber \\
  &=&\frac{d {\rm log} \omega_{log}}{dP} + (\frac{d {\rm log} f_1}{d\lambda}
            + \frac{d {\rm log} E(\lambda)}{d\lambda})\frac{d\lambda}{dP},
\label{eqn5}
\end{eqnarray}
where $f_1$ is the strong coupling correction and $E(\lambda)$ is
the exponential term in the Allen-Dynes equation.  The pressure
variation of the $M\omega_H^2$ denominator is challenging to
approximate without full calculations of the spectrum and
$\alpha^2F$. We have modeled the variation of the H spectrum by
assuming the three 3-fold $\Gamma$ point optic modes are
representative. The lower two of these modes are IR-active involving
H-S bond stretch and bond bending modes, the hardest frequency is a
silent mode with quadrupolar H motion with respect to S.

From calculations of these $\Gamma$ frequencies in the 240-270 GPa
range using the ELK code we calculate a positive but modest pressure
increase $d {\rm log} {\bar
  \omega}_H/dP$$\approx$1.9$\times$10$^{-3}$ GPa$^{-1}$.  From
Fig.~\ref{i2eta} we obtain $d {\rm
  log}\eta_H/dP$=3.5$\times$10$^{-3}$ GPa$^{-1}$, thus $\lambda$
decreases with pressure approximately as $d{\rm
  log}\lambda_H/dP$$\sim$-0.3$\times$10$^{-3}$/GPa. However, to our
precision this is indistinguishable from zero, so the pressure
derivative in Eq.~\ref{eqn5} reduces to the first term, the
frequency derivative.  The resulting prediction is a small increase
$dT_c/dP$=0.4 K/GPa.  This result disagrees in sign with Duan {\it
  et al.}, who quoted a smaller (in magnitude) negative value of
$dT_c/dP$=-0.12 K/GPa from direct calculation, however both numbers
are small compared to the large value of T$_c$ itself, so there is
no significant disagreement.

\section{Conclusions}

The report by Drozdov, Eremets, and Troyan of T$_c$ up to 190 K in
H$_n$S samples has breathed new life into the 50 year old
expectation of high T$_c$ in atomic H systems.  Both Li {\it et al.}
and Duan {\it et al.} had found that Eliashberg theory and linear
response results for electron-phonon coupling account for T$_c$ in
the 80-200 K range for H$_2$S and H$_3$S at high pressure, and the
analysis of Bernstein {\it et al} make $Im{\bar 3}m$ H$_3$S the
primary candidate to be this record-high temperature
superconductor. In this paper we have established that the coupling
of H vibrations increases strongly for pressures up to and even
beyond 210 GPa, and that 90+\% of the coupling arises from H
vibrations in this hydride that is stabilized by
hybridization\cite{bernstein} with S.  This picture is analogous to
the finding of the essential contribution of H in the superconductor
PdH at ambient pressure,\cite{PdH} and the broader picture of
Ashcroft\cite{ashcroft} of superconducting elemental H at high
pressure. The theoretical isotope shift of T$_c$ based on the
harmonic approximation is not in agreement with the experimental
result, suggesting substantial H anharmonicity will be necessary to
understand before the picture is complete.  Our picture, which
relies on coupling across the large Fermi surface, is at odds with
the hole superconductivity picture of Hirsch and
Marsiglio.\cite{hirsch}


\vskip 3mm

{\it Acknowledgments.} 

The authors acknowledge many insightful conversations on the theory
and application of GG theory with the late B. L. Gy\"{o}rffy, to
whom we dedicate this paper. We acknowledge discussions with
A. S. Botana, F. Gygi, and I. I. Mazin. M.J.M. was supported by the
Office of Naval Research through the US Naval Research Laboratory's
basic research program. W.E.P. was supported by NSF award
DMR-1207622-0. D.A.P. was supported by grant N00173-11-1-G002 from
the U.S. Naval Research Laboratory.


\begin{thebibliography}{10}

\bibitem{Drozdov}A. P. Drozdov, M. I. Eremets, and I. A. Troyan,
  arXiv:1412.0460.

\bibitem{Hg2223}A. Schilling. M. Cantoni, J. D. Guo, and H. R. Ott, Nature
  {\bf 363}, 6424 (1993).

\bibitem{Hg2223v2}C. W. Chu, L. Gao, F. Chen, Z. J. Huang, R. L. Meng,
  and Y. Y. Xue, Nature {\bf 365}, 323 (1993).

\bibitem{sch1}References to the experimental literature are given by
  J. S. Schilling, Physica C {\bf 460-462}, 182 (2007) and M.
  Debessai, J. J. Hamlin, and J. S. Schilling, Phys. Rev. B {\bf 78},
  064519 (2008). 

\bibitem{Li1}D. Kasinathan, J. Kune\v{s}, A. Lazicki, H. Rosner, 
  C. S. Yoo, R. T. Scalettar, and W. E. Pickett,  
  Phys. Rev. Lett. {\bf 96}, 047004 (2006).

\bibitem{Li2}D. Kasinathan, K. Koepernik, J. Kunes, H. Rosner, 
  and W. E. Pickett, Physica C {\bf 460-462}, 133-6 (2007).

\bibitem{DAPli}L. Shi and D. A. Papaconstantopoulos, Phys. Rev. B
  {\bf 73}, 184516 (2006).

\bibitem{calcium}Z. P. Yin, F. Gygi, and W. E. Pickett,  
  Phys. Rev. B {\bf 80}, 184515 (2009).

\bibitem{yttrium}Z. P. Yin, S. Y. Savrasov, and W. E. Pickett,  
Phys. Rev. B {\bf 74}, 094519 (2006).

\bibitem{CaY}S. Lei, D. A. Papaconstantopoulos, and M. J. Mehl,
Phys. Rev. B {\bf 75}, 024512 (2007).

\bibitem{silicon}N. Buckel and J. Wittig, Phys. Lett. {\it 17}, 187 (1965).

\bibitem{ashcroft}N. W. Ashcroft, Phys. Rev. Lett. {\bf 21}, 1748 (1968).

\bibitem{ashcroft2}N. W. Ashcroft, Phys. Rev. Lett. {\bf 92}, 187002 (2004).

\bibitem{Li}Y. Li {\it et al.}, J. Chem. Phys. {\bf 140}, 174712 (2014).

\bibitem{alldyn}P. B. Allen and R. C. Dynes, Phys. Rev. B {\bf 12}, 905 (1975).

\bibitem{Duan}D. Duan {\it et al.}, Sci. Rep. {\bf 4}, 6968 (2014).

\bibitem{GG}G. D. Gaspari and B. L. Gy\"orffy, Phys. Rev. Lett. {\bf 28}, 801 (1972).

\bibitem{hydrogen}D. A. Papaconstantopoulos and B. M. Klein, Ferroelectrics
  {\bf 16}, 307 (1977); D. A. Papaconstantopoulos {\em et al.},
  Phys. Rev. B {\bf 15} 4221 (1977).

\bibitem{Hopfield}J. J. Hopfield, Phys. Rev. {\bf 186}, 443 (1969).

\bibitem{bernstein}N. Bernstein, C. S. Hellberg, M. D. Johannes, I. I. Mazin,
  and M. J. Mehl, Phys. Rev. B {\bf 91}, 060511(R) (2015).

\bibitem{SSW}D. J. Scalapino, J. R. Schrieffer, and J. W. Wilkins,
  Phys. Rev. {\bf 148}, 263 (1966).

\bibitem{klein76} B. M. Klein and D. A. Papaconstantopoulos,
  J. Phys. F: Metal Phys. {\bf 6}, 1135 (1976).

\bibitem{wep}W. E. Pickett, Physica B+C {\bf 111B}, 1 (1981).

\bibitem{nrllapw}The NRL LAPW code, originally developed by
  H. Krakauer and D. J. Singh, was used with Hedin-Lundqvist
  exchange-correlation. DOS results were generated from 285 k points
  in the irreducible Brillouin zone with the tetrahedron
  method. Total energies were fit to the Birch equation to obtain
  the P(V) equation of state.

\bibitem{elk} http://elk.sourceforge.net

\bibitem{PdH}D. A. Papaconstantopoulos and B. M. Klein,
  Phys. Rev. Lett.  {\bf 35}, 110 (1975); B. M. Klein {\em et al.},
  Phys. Rev. Lett {\bf 39}, 574 (1977); D. A. Papaconstantopoulos,
  B. M. Klein, E. N. Economou and L. L. Boyer Phys. Rev. B {\bf 17},
  141 (1978).

\bibitem{hirsch}J. E. Hirsch and F. Marsiglio, arXiv:1412.6251.

\end{thebibliography}
\end{document}